\documentclass[12pt]{article}
\usepackage{geometry}             
\usepackage{graphicx}
\usepackage{amssymb}
\usepackage{amsmath}
\usepackage{epstopdf}

\usepackage[hyperindex=true,
          pdfstartview=FitH,
          bookmarksnumbered=true,
          bookmarksopen=true,
          citecolor=blue,
          linkcolor=blue,
          colorlinks=true,
          unicode]{hyperref}

\begin{document}

\title{Accelerating vacua in Gauss-Bonnet gravity}
\author{Wei Xu, Kun Meng and Liu Zhao\thanks{email: {\it lzhao@nankai.edu.cn}}\\
School of Physics, Nankai university, Tianjin 300071, China}
\date{}                    
\maketitle

\begin{abstract}
Accelerating vacua with maximally symmetric, but not necessarily spherical, sections for
Einstein and Gauss-Bonnet gravities in generic dimensions are obtained. The 
acceleration parameter has the effect of shifting the cosmological constants in Einstein 
gravity, whereas in Gauss-Bonnet gravity the effective cosmological constants remain the 
same in the presence of acceleration as in the case without acceleration.
\end{abstract}

\section{Introduction}

In a recent note \cite{Zhao:2011p693}, one of the authors found that, in $n$-dimensions, adding a conformal factor
\begin{align}
\omega^{2}(r,\theta_{1})=\frac{1}{(1-\alpha r \cos\theta_{1})^{2}} \label{om}
\end{align}
in front of the standard de Sitter, Minkowski and/or anti-de Sitter metrics still makes 
Einstein vacua, with cosmological constant appropriately modified. The result 
can be summarized as follows. Let
\begin{align}
ds_{\Lambda}^{2}&=-f(r) dt^{2}
+f(r)^{-1}dr^{2}+r^{2}d\Omega_{n-2}^{2} \label{dsl}
\end{align}
with
\begin{align}
f(r)&=1-\frac{2\Lambda}{(n-1)(n-2)}r^{2} \label{f1}
\end{align}
representing an Einstein vacuum solution with cosmological constant $\Lambda$, where
$d\Omega_{n-2}^{2}$ represents the line element of an $(n-2)$-dimensional
sphere spanned by the angular coordinates $(\theta_{1},\theta_{2},...,
\theta_{n-2}=\phi)$. Then, the conformally transformed metric
\begin{align}
ds_{\Lambda,\alpha}^{2}=\omega^{2}(r,\theta_{1})ds_{\Lambda}^{2} \label{conf1}
\end{align}
is also an Einstein vacuum with cosmological constant
\begin{align}
\Lambda_{\alpha}=\Lambda -\frac{(n-1)(n-2)\alpha^{2}}{2}.  \label{laph}
\end{align}
For $\alpha\ne 0$, the cosmological constant $\Lambda_{\alpha}$ in the 
presence of the conformal factor is different from the original cosmological constant
$\Lambda$. That the metric (\ref{conf1}) represents an accelerating vacua was known
before the work \cite{Zhao:2011p693} for $n=3$ \cite{Astorino:2011p1101}, 
$n=4$ \cite{Dias:2003p2357,Dias:2002p2471} and $n=5$ \cite{Xu:2010p5603}.

Each of the constant $r =C \ne 0$ hyper surface in (\ref{dsl}) represents a conicoid (or part of a conicoid) with eccentricity $\alpha C$, and one of the foci of the 
conicoid is located at the spacial origin $r=0$.
The physical interpretation of the parameter $\alpha$ is clear: it is the magnitude of the
proper acceleration of the static observer located at the specified focus at 
the origin.

In another recent work \cite{Zhao:2011p1230}, 
we found some accelerating vacua for Gauss-Bonnet gravity in 5 and 6 dimensions, 
with angular sections being either spherical or hyperbolic. For hyperbolic cases, the 
conformal factor in the metric is different from the one given in (\ref{om}).

Physically, accelerating vacua are of interests because they are the vacua
to be perceived by observers under relative accelerations with respect to the free-falling
observers. In the study of gravitational theories, equivalence principle is often used to 
cancel out local gravitational effect by changing into the free-falling frames. However, 
in some cases, it might be necessary to make use of the equivalence principle in the 
converse way. Imagine that one day our human beings might be able to make 
intergalactic voyages. In such 
circumstances, it will be inevitable to encounter the case in 
which the observers undergo relative acceleration with respect to the free-falling
observers, and equivalence principle predicts that the observed structure 
of spacetime  for observers in the spaceship should be different from that for the 
free-falling observers. Though higher dimensional accelerating vacua
are not directly related to such practical applications, they still constitute an essential 
part of gravity theories in the corresponding spacetime dimension and thus should be of
academic interests.

The present work is aimed at extending the previous results to more general settings, e.g. 
allowing $d\Omega_{n-2}^{2}$ to be replaced by the metric of a maximally 
symmetric subspace  and/or extending the results to 
higher curvature gravity theories such as Gauss-Bonnet gravity. 
We shall accomplish our 
goal via two steps: accelerating vacua for Einstein gravity will 
be considered in Section 2, and extensions to Gauss-Bonnet gravity will be presented in
Section 3. Proper accelerations of the vacua are analyzed in Section 4. And finally we 
give the conclusion in Section 5.

\section{Accelerating vacua in Einstein gravity}

It is well known that a generalization of the metric (\ref{dsl}) to the case with maximally symmetric sections exists, the metric is given by\footnote{The metric (\ref{dsl2}) with 
$f_{k}(r)$ given in
(\ref{f2}) is the zero mass zero charge limit of the so-called topological 
AdS black hole solution found in 
\cite{Mann:1996p968,Vanzo:1997p1331,Birmingham:1998p1113}. With nonzero,
positive mass, the metric in 
\cite{Mann:1996p968,Vanzo:1997p1331,Birmingham:1998p1113} was called topological 
AdS black hole because $k=0,-1$ are allowed only for $\Lambda<0$. 
For $\Lambda\ge 0$ and $k=0,-1$, the corresponding 
solution does not allow for the existence of a horizon due to 
Hawking's black hole topology theorem \cite{Hawking} and its cousin in higher 
dimensions \cite{Galloway}. The restriction 
$\Lambda<0$ for $k=0,-1$ holds even in the zero mass limit, because we need a region 
in the spacetime in which $f_{k}(r)>0$ in order to interpret $t$ as a timelike coordinate.}
\begin{align}
ds_{\Lambda,k}^{2}&=-f_{k}(r) dt^{2}
+f_{k}(r)^{-1}dr^{2}+r^{2}d\Sigma_{n-2,k}^{2},\label{dsl2}
\end{align}
where
\begin{align}
f_{k}(r)&=k-\frac{2\Lambda}{(n-1)(n-2)}r^{2}. \label{f2}
\end{align}
Here $k=0,\pm 1$ and $d\Sigma_{n-2,k}^{2}$ 
represents the metric of an $(n-2)$-dimensional maximally symmetric manifold and can 
be written explicitly as
\begin{align}
d\Sigma_{n-2,k}^{2}&=d\theta_{1}^{2}
+ \rho_{k}^{2}(\theta_{1}) d\Omega_{n-3}^{2},
\label{dsigma}\\
\rho_{k}(\theta_{1})&=\lim_{\kappa\rightarrow k}\frac{\sin\left(\sqrt{\kappa}\,\theta_{1}\right)}{\sqrt{\kappa}}
=\left\{
\begin{array}{ll}
\sin\theta_{1} & (k=+1) \cr
\theta_{1} & (k=0)\cr
\sinh\theta_{1} & (k=-1)
\end{array}
\right.  \label{rho}
\end{align}
where $d\Omega_{n-3}^{2}$ is the line element of an $(n-3)$-sphere. The choice 
$k=1$ corresponds to the spherically symmetric case 
(\ref{dsl}) and $k=0, -1$ correspond to flat and hyperbolic cases, respectively. 
The metric (\ref{dsl2}) with insertions (\ref{f2})-(\ref{rho}) solves the 
vacuum Einstein equation
\begin{align}
R_{MN}-\frac{1}{2}g_{MN}R+\Lambda g_{MN}=0.  \label{ve}
\end{align}
We can rewrite the function $f_{k}(r)$ in a more familiar form as
\begin{align}
f_{k}(r)&=k-\epsilon\frac{r^{2}}{\ell^{2}}, \label{fell}
\end{align}
where $\ell$ represents the (A)dS radii and $\epsilon$ is the sign of $\Lambda$,\begin{align}
\frac{1}{\ell^{2}}=\frac{2|\Lambda|}{(n-1)(n-2)},
\qquad \epsilon=\mathrm{sign}(\Lambda).
\end{align}
One can read off the cosmological constant $\Lambda$ from the metric function (\ref{fell})
using
\begin{align}
\Lambda = \frac{(n-1)(n-2)\epsilon}{2\ell^{2}}. \label{lambdaell}
\end{align}

It is natural to ask what the $k=0,-1$ analogues of (\ref{conf1}) are. After some tedious 
tensor algebra we found the following to be the right answer:
\begin{align}
ds_{\Lambda,k,\alpha}^{2}=\omega_{k}^{2}(r,\theta_{1})ds_{\Lambda,k}^{2},  
\label{dlk2}
\end{align}
where
\begin{align}
\omega_{k}^{2}(r,\theta_{1})=\frac{1}{(1-\alpha r \sigma_{k}(\theta_{1}))^{2}},
\label{omega}
\end{align}
and
\begin{align}
\sigma_{k}(\theta_{1})=\cos\left(\sqrt{k}\,\theta_{1}\right)
=\left\{
\begin{array}{ll}
\cos\theta_{1} & (k=+1) \cr
1 & (k=0)\cr
\cosh\theta_{1} & (k=-1)
\end{array}
\right.
\label{sigmauni}
\end{align}
It can be easily checked 
that the metric (\ref{dlk2}) obeys the vacuum Einstein equation (\ref{ve}) with the 
cosmological constant $\Lambda$ replaced by
\begin{align}
\Lambda_{\alpha}=\Lambda-\frac{(n-1)(n-2)k\alpha^{2}}{2}. \label{lambdalambda}
\end{align}
For $k=+1$, this reproduces the known result (\ref{laph}). Notice that for $k=\pm 1$, a 
nonzero cosmological constant $\Lambda_{\alpha}$ can be produced by the presence 
of the parameter $\alpha$ even if we start from $\Lambda=0$.

\section{Accelerating vacua in Gauss-Bonnet gravity}

Gauss-Bonnet gravity has the equation of motion
\begin{align}
R_{\mu\nu}-\frac{1}{2}g_{\mu\nu}R + \Lambda g_{\mu\nu} + \xi H_{\mu\nu}
=0, \label{eqm1}
\end{align}
where
\begin{align}
H_{\mu\nu}=2\left(
R_{\mu\lambda\rho\sigma}R_{\nu}{}^{\lambda\rho\sigma}
-2R_{\mu\rho\nu\sigma}R^{\rho\sigma}-2R_{\mu\sigma}R_{\nu}{}^{\sigma}
+RR_{\mu\nu}\right)
-\frac{1}{2}\mathcal{L}_{\mathrm{GB}} g_{\mu\nu}. \label{eqm2}
\end{align}
This is the consequence of variations of the action (dropping an appropriate boundary 
term) of Gauss-Bonnet gravity in $n$-dimensions
\begin{align*}
&I=\frac{1}{16\pi G}\int \mathrm{d}^n x
\sqrt{-g}\left[R-2\Lambda+\xi \mathcal{L}_{\mathrm{GB}}
\right],\\
&\mathcal{L}_{\mathrm{GB}}=
R_{\mu\nu\gamma\delta}R^{\mu\nu\gamma\delta}
-4R_{\mu\nu}R^{\mu\nu}+R^2, 
\end{align*}
where $G$ is the $n$-dimensional Newton constant, $\Lambda$ is a  
cosmological constant and $\xi$ is the Gauss-Bonnet parameter.

The standard vacua of the form (\ref{dsl2}) for Gauss-Bonnet gravity 
are given in terms of the metric function $f_{k}(r)$ as
\begin{align}
(ds_{\mathrm{GB}})_{\Lambda,k}^{2}&=-f_{k}(r) dt^{2}
+f_{k}(r)^{-1}dr^{2}+r^{2}d\Sigma_{n-2,k}^{2},\label{dsgb}
\end{align}
\begin{align}
f_{k}(r)&=k+\frac{1}{2(n-3)(n-4)\xi}(1+\delta\sqrt{a_{n}}\,)r^{2},  \label{fgb1}
\end{align}
where
\[
a_{n}=1+ \frac{8(n-3)(n-4)\xi \Lambda}{(n-1)(n-2)}
\]
and $\delta=\pm 1$ indicate that there are two branches for the vacua which is the 
consequence of the $R^{2}$ form of the action. These 
vacua are obtained from the Gauss-Bonnet black hole solutions in the zero mass limit. 
The $k=+1$ version of these vacua was first found by Boulware and Deser in 
\cite{Boulware:1985p1145}, and the general form (\ref{fgb1}) can be inferred from
\cite{Cai:2001p6111, Dehghani:2004p869}. The same solution can also be obtained 
from the works \cite{Cvetic:2001p1725,Nojiri:2002p1316} on general curvature 
squared gravity theory by taking the Gauss-Bonnet limit.

The function $f_{k}(r)$ defined in 
(\ref{fgb1}) can also be written in the form (\ref{fell}), but now with $\ell$ and $
\epsilon$ given as follows:
\begin{align}
\frac{1}{\ell^{2}}=\frac{\left|1+\delta\sqrt{a_{n}}\right|}{2(n-3)(n-4)\xi},
\qquad \epsilon=-\mathrm{sign}(1+\delta\sqrt{a_{n}}).
\end{align}
This indicates that the solutions (\ref{dsgb}) 
are also Einstein vacua, with effective cosmological constants
\begin{align}
\Lambda_{\mathrm{eff}}&=
-\frac{(n-1)(n-2)}{4(n-3)(n-4)\xi}
\left(1+\delta\sqrt{a_{n}}\,\right),   \label{leff}
\end{align}
thanks to the relation (\ref{lambdaell}).

Now we would like to ask whether the accelerating vacua of the form similar to (\ref{dlk2}) exists for Gauss-Bonnet gravity. A direct substitution yields that
the ansatz
\begin{align*}
(ds_{\mathrm{GB}})_{\Lambda,\alpha,k}^{2}=\omega_{k}^{2}(r,\theta_{1})
(ds_{\mathrm{GB}})_{\Lambda,k}^{2} 
\end{align*}
with $(ds_{\mathrm{GB}})_{\Lambda,k}^{2}$ given as in (\ref{dsgb})-(\ref{fgb1}) 
fails to solve the Gauss-Bonnet field equations (\ref{eqm1})-(\ref{eqm2}) unless
$k=0$ or $\alpha=0$. However, 
one can check that the following metrics indeed solve the field equation  
for Gauss-Bonnet gravity:
\begin{align}
(ds_{\mathrm{GB}})_{\Lambda,k,\alpha}^{2}
&=\omega_{k}^{2}(r,\theta_{1}) ds_{k,\alpha}^{2}, \label{lka}\\
ds_{k,\alpha}^{2} &=-f_{k,\alpha}(r) dt^{2} +f_{k,\alpha}(r)^{-1}dr^{2} + 
r^{2}d\Sigma_{n-2,k}^{2},\\
f_{k,\alpha}(r)&=k-\left(k \alpha^{2}- \frac{1}{2(n-3)(n-4)\xi }
\left(1+\delta\sqrt{a_{n}}\,\right)\right)r^{2}. \label{fkalph}
\end{align}
Here, again, $\omega_{k}^{2}(r,\theta_{1})$ is given by (\ref{omega}).  
For $k=\pm 1$ and $n=5,6$, the above solution reproduces the results of
\cite{Zhao:2011p1230}. We can also rewrite (\ref{fkalph}) as
\begin{align}
f_{k,\alpha}(r)&=k-\epsilon\frac{r^{2}}{\ell^{2}}, \label{fellalph}
\end{align}
but now with
\begin{align*}
\frac{1}{\ell^{2}}&=\left|k \alpha^{2}-\frac{1}{2(n-3)(n-4)\xi }
\left(1+\delta\sqrt{a_{n}}\,\right)\right|,\\
\epsilon&=\mathrm{sign}\left(k \alpha^{2}- \frac{1}{2(n-3)(n-4)\xi }
\left(1+\delta\sqrt{a_{n}}\,\right)\right).
\end{align*}
This shows that the metrics represented by $ds_{k,\alpha}^{2}$ 
are Einstein vacua (however, \textit{not} Gauss-Bonnet vacua, which explains the 
absence of  the suffix GB in the  notation), with the 
corresponding cosmological constants (due to (\ref{lambdaell}))
\begin{align*}
\Lambda_{k,\alpha}=\frac{(n-1)(n-2)}{2}
\left(k\alpha^{2} - \frac{1}{2(n-3)(n-4)\xi }
\left(1+\delta\sqrt{a_{n}}\,\right)\right).
\end{align*}
In the $\alpha\rightarrow 0$ limit, the metrics $ds_{k,\alpha}^{2}$ will approach
the Gauss-Bonnet vacua (\ref{dsgb}).

The Gauss-Bonnet vacua (\ref{lka}) are also Einstein vacua,  
with cosmological constants given as 
\begin{align*}
\Lambda_{\alpha} &= \Lambda_{k,\alpha}-\frac{(n-1)(n-2)k\alpha^{2}}{2}\\
&=-\frac{(n-1)(n-2)}{4(n-3)(n-4)\xi}
\left(1+\delta\sqrt{a_{n}}\,\right),
\end{align*}
which is identical to the effective cosmological constant $\Lambda_{\mathrm{eff}}$ 
(eq.(\ref{leff})) for the non accelerating Gauss-Bonnet vacua (\ref{dsgb}).
We stress that the effective cosmological constants for the accelerating vacua of 
Gauss-Bonnet gravity depend neither on $k$ nor 
on the parameter $\alpha$. This is in sharp contrast to the case of pure 
Einstein gravity.

\section{Proper accelerations of the vacua}

So far we have been referring to the spacetimes with conformal factor 
$\omega_{k}^{2}(r,\theta_{1})$ with the term ``accelerating vacua'' 
without justification for the appropriateness of this terminology. 
Now we are in a position to settle this problem.

Both the vacua (\ref{dlk2}) of Section 2 and (\ref{fkalph}) of Section 3 
can be written in a unified way as
\begin{align}
ds^{2}=\omega_{k}^{2}(r,\theta_{1})\left[
-\left(k-\epsilon\frac{r^{2}}{\ell^{2}}\right)dt^{2}
+\left(k-\epsilon\frac{r^{2}}{\ell^{2}}\right)^{-1}dr^{2}
+r^{2}d\Sigma_{n-2,k}^{2}\right]. \label{xmu}
\end{align}
This allows us to look at the static observers 
\[
x^{\mu}=\left(\tau\frac{1-\alpha r \sigma_{k}(\theta_{1})}
{\sqrt{k-\epsilon r^{2}/\ell^{2}}},r, \theta_{1}..., \theta_{n-2}\right)
\]
with proper velocity
\begin{align*}
u^{\mu}=\left(\frac{1-\alpha r \sigma_{k}(\theta_{1})}
{\sqrt{k-\epsilon r^{2}/\ell^{2}}},0,...,0\right),
\end{align*}
where $\tau$ is the proper time. Note that in writing (\ref{xmu}) we have assumed 
that the observer lives in a region of the spacetime in which 
$k-\epsilon r^{2}/\ell^{2}>0$, otherwise the observer will not be static. This requires,
in particular, $\epsilon=-1$ for $k=0,-1$. Straightforward calculations show that the 
proper acceleration $a^{\mu}=u^{\nu}\nabla_{\nu}u^{\mu}$ has only two nonzero 
components, i.e.
\begin{align*}
&a^{r}=\frac{1}{\ell^{2}}(1-\alpha r \sigma_{k}(\theta_{1}))
(\epsilon r+k \ell^{2}\alpha \sigma_{k}(\theta_{1})),\\
&a^{\theta_{1}}=(1-\alpha r \sigma_{k}(\theta_{1})) \frac{\alpha}{r}
\frac{d\sigma_{k}(\theta_{1})}{d\theta_{1}}.
\end{align*}
Contracting $a^{\mu}$ with itself, we get 
\begin{align}
a^{\mu}a_{\mu}=
\alpha^{2}\left(\frac{d\sigma_{k}(\theta_{1})}{d\theta_{1}}\right)^{2}
+\frac{1}{\ell^{2}}\frac{(\epsilon r/\ell +k\alpha\sigma_{k}(\theta_{1}))^{2}}
{k-\epsilon r^{2}/\ell^{2}}. \label{amag}
\end{align}
For $k=\pm 1$, the squared norm $a^{\mu}a_{\mu}$ at the origin reads
\begin{align*}
a^{\mu}a_{\mu}=k \alpha^{2},
\end{align*}
which justifies our statement that $\alpha$ is proportional to the magnitude of the 
proper acceleration of the observer at the origin. However, we should note that for 
$(\epsilon=-1, k=0)$, the point $r=0$ is a singularity of the metric, so it is meaningless 
to talk about the proper acceleration at the origin in this case. Beyond the singularity
at $r=0$, the proper acceleration for $(\epsilon=-1, k=0)$ is uniform
and independent of $\alpha$:
\begin{align*}
a^{\mu}a_{\mu}=\frac{1}{\ell^{2}}.
\end{align*}

Let us remark that, since $a^{t}=0$, the proper acceleration $a^{\mu}$ is spacelike 
when $t$ is timelike. So we ought to have $a^{\mu}a_{\mu}>0$ in the static region.
This is indeed the case as can be seen from (\ref{amag}), where the static region is 
determined by the condition $k-\epsilon r^{2}/\ell^{2}>0$. Notice that for $k=-1, 
\epsilon=-1$, the condition for the static region is reduced into $r>\ell$, 
the origin $r=0$ is clearly not in the static region. This explains why 
$a^{\mu}a_{\mu}=-\alpha^{2}<0$ at $r=0$ for $k=-1$. Notice also that, for the two 
distinguished cases $(\epsilon=+1, k=+1)$ and $(\epsilon=-1, k=-1)$, 
$a^{\mu}a_{\mu}$ 
becomes infinity at $r=\ell$. This implies that the $r=\ell$ hyper surfaces in these two 
cases are accelerating horizons. The difference between the two cases lies in that, for 
$(\epsilon=+1, k=+1)$, the condition for the static region is $r<\ell$, so the acceleration 
horizon is like a cosmological horizon (i.e. static observers are located inside the 
horizon); while for $(\epsilon=-1, k=-1)$, the condition 
for the static region is $r>\ell$, and the acceleration horizon is like a black hole 
horizon (static observers are located outside the horizon)\footnote{
For $k=0,-1$, the hyper surface with metric $d\Sigma_{n-2,k}^{2}$ are noncompact,
so, at first sight, it is meaningless to talk about the inside and outside of the horizon.
However, one can consider these cases modulo a discrete symmetry group, making the
quotient compact. In this sense one can indeed talk about the inside and outside of the
horizon for $k=-1$.}.

\section{Conclusions}

We have thus addressed the problem of finding accelerating vacua for Einstein and
Gauss-Bonnet gravities with maximally symmetric, but not necessarily 
spherical, sections. For Einstein gravity, the accelerating vacua can be 
obtained by conformally transforming some given, non accelerating, Einstein vacua
with some prescribed cosmological constants. Then, the accelerating vacua will 
correspond to a different cosmological constant, i.e. acceleration has the effect of shifting 
cosmological constant for Einstein gravity.

For Gauss-Bonnet gravity, the accelerating vacua cannot be obtained by conformally 
transforming non accelerating Gauss-Bonnet vacua. Instead, they 
can be obtained by conformally transforming some non accelerating 
Einstein vacua, and the effective cosmological constants remain the same as those
for non accelerating Gauss-Bonnet vacua. 

It should be remarked that the solutions we obtained here are only accelerating Einstein vacua without black holes. It is more tempting to get the accelerating black hole 
solutions but this seems to be extremely difficult. The physical reason for why the 
accelerating black holes are difficult to find is because that 
it requires tremendous energy to accelerate black holes and hence the corresponding 
metrics would in general not vacua. However, just what kind of matter source can 
provide the energy necessary to accelerate black holes in Einstein and Gauss-Bonnet 
gravities remain an open problem, the only exceptions are the 4-dimensional 
C-metric black hole solutions found long ago by Levi-Civita 
(see \cite{Dias:2003p2357,Dias:2002p2471} for 
detailed description for versions with cosmological constant) for Einstein gravity, in
which conical singularities on the horizons play the role of source for the acceleration.

\section*{Acknowledgment} 

This work is supported by the National  Natural Science Foundation of 
China (NSFC) through grant No.10875059. 


\providecommand{\href}[2]{#2}\begingroup\raggedright\endgroup

\end{document}